 \documentclass[
  ,final            
  ,numberedheadings 
  ]
  {aipproc}

\layoutstyle{8x11double}

\begin{document}

\title 
      [Evolution of massive stars at very low Z]
      {Evolution of massive stars at very low metallicity including rotation and binary interactions}


\author{S.-C Yoon}{
  address={Department of Astronomy and Astrophysics, University of California, Santa Cruz, CA95064, USA},
  email={scyoon@ucolick.org},
}

\iftrue
\author{M. Cantiello}{
  address={Astronomical Institute, Utrecht University, Princetonplein 5, 3584 CC, Utrecht, The Netherlands},
  email={m.cantiello@astro.uu.nl},
}

\author{N. Langer}{
  address={Astronomical Institute, Utrecht University,  Princetonplein 5, 3584 CC, Utrecht, The Netherlands},
  email={n.langer@astro.uu.nl},
}
\fi

\copyrightyear  {2007}

\begin{abstract}
We discuss recent models on the evolution of massive stars at very low metallicity including the effects
of rotation, magnetic fields and binarity. 
Very metal poor stars lose very little mass and angular momentum during the main sequence evolution, 
and rotation plays a dominant role in their evolution. 
In rapidly rotating massive stars, the rotationally induced mixing time scale can be even 
shorter than the nuclear time scale  throughout the main sequence. The consequent 
quasi-chemically homogeneous evolution greatly differs from the standard massive star evolution 
that leads to formation of red giants with strong chemical stratification. 
Interesting outcomes of such a new mode of evolution include the
formation of rapidly rotating massive Wolf-Rayet stars that emit large amounts of ionizing photons, 
the formation of a long gamma-ray bursts and a hypernovae, and the production of large amounts of primary nitrogen. 
We show that binary interactions can further enhance the effects of rotation, as mass accretion in a close binary
spins up the secondary. 
\end{abstract}

\date{\today}

\maketitle

\section{Introduction}

Massive stars affect the evolution of the early universe in a number of ways. 
They are believed to be the main source of reionizing photons at high redshift, 
and their explosions provide large amounts of energy into the surrounding medium.
Most elements heavier than helium begin to exist in the early universe
due to nucleosynthesis in massive stars.
The history of star formation and the evolution of galaxies 
in the early universe are therefore closely related to such feedback from massive stars.
This has motivated many theoretical studies on the evolution of massive stars 
of the first and second generations, which are characterized by zero/low metallicity.

Mass loss due to stellar winds from massive stars is mainly driven by metal lines 
(Castor, Abbott \& Klein~1975; Pauldrach, Puls \& Kudritzki R.P.~1986; Vink, de Koter \& Lamers~2001).
As a result, metal poor massive stars are expected to lose only a small amount of mass and angular momentum
and to be kept rotating rapidly, during their life times (see Ekstr\"om and Meynet, however,  in this volume). 
Rotation -- one of the primary factors to determine the evolution of massive stars --
may thus play an even more important role in the evolution of massive stars in the early universe, compared
to the case of galactic metal-rich massive stars. 
For instance, the high ratio of nitrogen to carbon abundance observed in extremely 
metal poor stars may be
related to chemical mixing due to rotationally induced instabilities in massive stars
in the early universe (Chiappini et al.~2006).
Stellar cores could also more easily retain a large amount of angular momentum at lower metallicity, 
resulting in the abundant production of energetic supernovae and gamma-ray bursts 
(Yoon \& Langer~2005; Woosley \& Heger~2006; Yoon, Langer \& Norman~2006).

Detailed numerical simulations of the evolution of massive stars including the effects of rotation involve many 
uncertain physical processes such as the transport of angular momentum and chemical species due to 
rotationally induced instabilities.
Studies by Langer et al. (1999), Heger, Langer \& Woosley (2000) and Hirschi, Meynet \& Maeder (2004)
indicate that their adopted angular momentum transport mechanisms (Eddington Sweet circulations, 
shear instability and baroclinic instability) are too inefficient to explain
the observed spin rates of white dwarfs and young neutron stars: 
their models predict one or two orders of magnitude higher spin rates than the observed values.
More recent models adopting the prescription by Spruit (2000) for the Tayler-Spruit dynamo
in differentially rotating radiative layers are shown to be more consistent with observations in terms
of the spin rates of stellar remnants (Heger, Woosley \& Spruit~2005; Suijs et al. in prep.). 
The validity of the Talyer-Spruit dynamo has recently been challenged
by several authors (Denissenkov, Pavel, \& Pinsonneault~2007; Zahn, Brun \& Mathis, S.~2007) 
but it seems clear that an efficient braking mechanism that is comparable to
what Spruit suggests is needed to explain observations.

Here we present recent models
of both single and binary massive stars at very low metallicity 
including the Tayler-Spruit dynamo as well as other effects of rotation
such as rotationally induced chemical mixing, enhanced mass loss near the break-up velocity
and tidal interactions. We suggest that rotation could lead to 
very different types of massive star evolution at very low metallicity, 
compared to the case of metal rich stars. Implications
for supernovae and gamma-ray bursts, and massive star feedback in the early universe are briefly discussed.  

\section{Evolution of single stars at very low metallicity}

\begin{figure}
\resizebox{0.98\hsize}{!}{\includegraphics{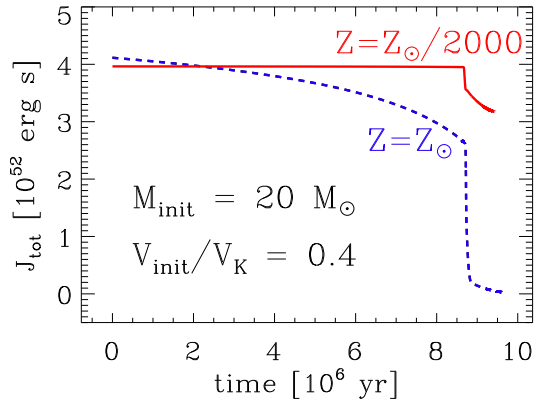}}
\caption{Total angular momentum in $20~\mathrm{M_\odot}$ models as a function
of evolutionary time. The solid line and the dashed lines give the results with $Z = 10^{-5}$ and $0.02$ respectively.}
\label{figjtot}
\end{figure}

\begin{figure}
\resizebox{0.98\hsize}{!}{\includegraphics{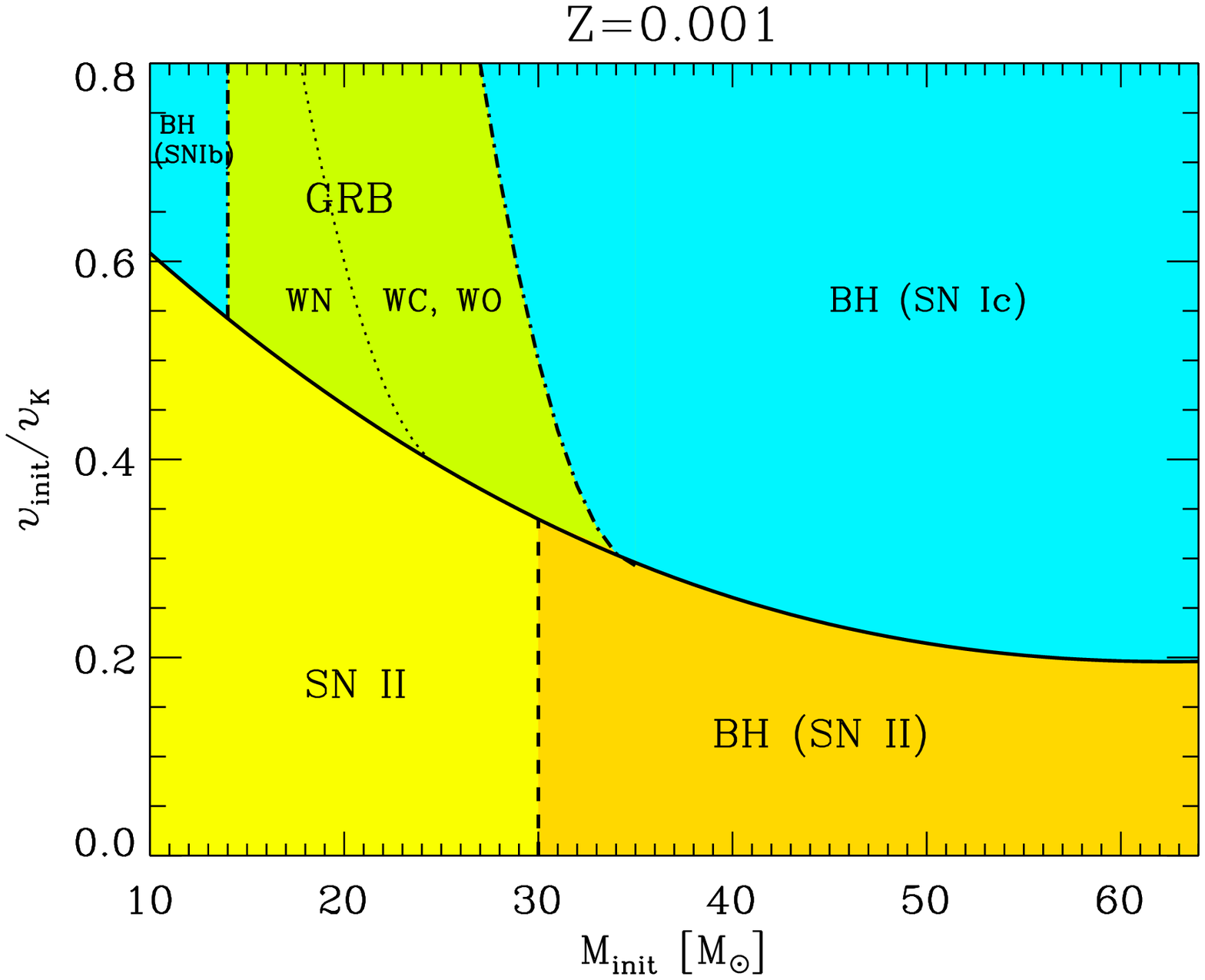}}
\resizebox{0.98\hsize}{!}{\includegraphics{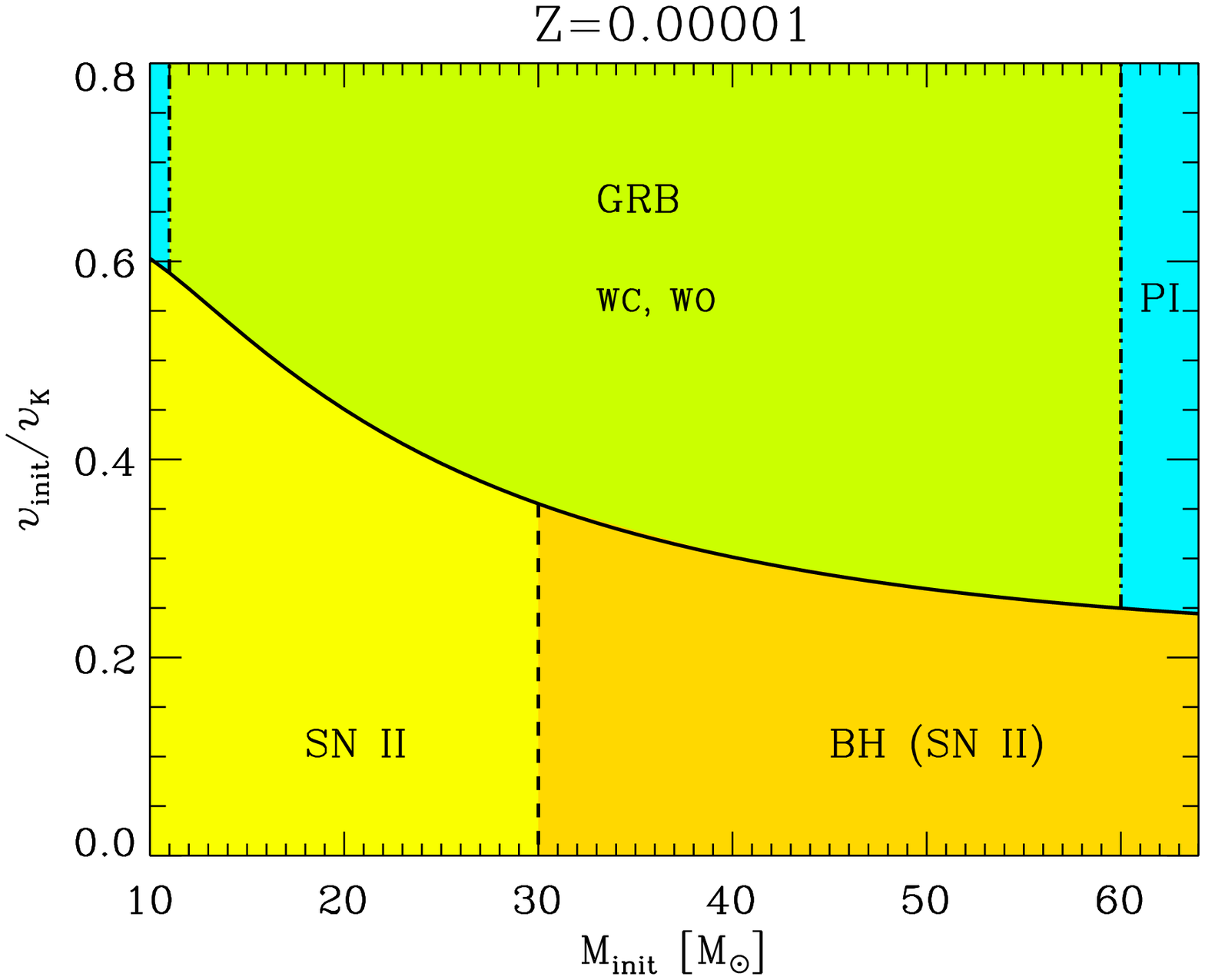}}
\caption{
Final fate of our rotating massive star models at two different metallicities ($Z=$  0.001
\& 0.00001), in the plane of initial mass and  initial fraction of the Keplerian value of the equatorial rotational velocity.
To both sides of the GRB production region for $Z=0.001$, black holes are
expected to form inside WR stars, but the core spin is insufficient to allow GRB production.
For $Z$ = 0.00001, the pair-instability might occur to the right side of the
GRB production region (see Heger et al 2003), although the rapid rotation may shift
the pair instability region to larger masses.
The dashed line in the region of non-homogeneous evolution
separates Type II supernovae (SN II; left) and black hole (BH; right) formation,
where the minimum mass for BH formation is simply assumed to be $30~\mathrm{M_\odot}$.
From Yoon, Langer \& Norman~(2006). 
}
\label{fig:fate}
\end{figure}

It is well known that many massive stars in our Galaxy as well as in the Small/Large Magellanic Cloud
are rapid rotators (e.g. Maeder \& Meynet~2000; Mokiem et al.~2006; Mokiem et al.~2007). 
One of the most important effects of rotation on the evolution of massive stars is
rotationally induced mixing of chemical elements (Maeder \& Meynet~2000; Heger \& Langer~2000).
In magnetic models, stars tend to rotate rigidly due to magnetic torques,  
and thus it is Eddington-Sweet circulations
that mainly induce chemical mixing inside stars (Maeder \& Meynet~2005; Heger, Woosley \& Spruit 2005; Yoon \& Langer 2005), 
rather than the shear instability that is important in non-magnetic models 
(Maeder \& Meynet 2000; Heger, Langer \& Woosley 2000; Meynet \& Maeder~2002). 
Efficient chemical mixing
across the boundary between the core and the envelope in a star by Eddington-Sweet circulations 
is usually prohibited due to the chemical gradient built up by nuclear burning in the convective core
(Zahn 1992). 
If a star is born with a sufficiently large amount of angular momentum, however, 
mixing could occur on a shorter time scale than the nuclear burning time scale. 
I.e.,  the star becomes chemically homogeneous due to efficient chemical mixing by Eddington Sweet circulations
and the classical core-envelope structure is not developed (Maeder 1987; Langer 1992). 
Such a star would be gradually transformed 
into a rapidly rotating massive helium star (i.e., WR star) by the end of the core hydrogen burning. 
This mode of evolution is supposed to be particularly important for metal poor massive stars. 
At high metallicity, strong winds significantly spin down a massive star early on the main sequence, 
rendering rotationally induced mixing too inefficient to lead to the chemically homogeneous evolution 
(see Fig.~\ref{figjtot}).

In Fig.~\ref{fig:fate}, we summarize how the evolution and final fate of massive stars differ according to  
the initial mass and rotational velocity, at two different metallicities, based on 
our stellar evolution models (Yoon, Langer \& Norman~2006). 
Slowly rotating stars follow the classical evolutionary path such that
they evolve redwards on the HR diagram, and end their life as a (super) giant
with a thick hydrogen envelope. In this case, the core is strongly braked down during the giant phase
by the heavy envelope, resulting in a specific angular momentum of the order of $10^{14}~\mathrm{cm^2~s^{-1}}$
in the pre-collapsing core. Type II supernovae are the expected outcome, leaving millisecond pulsars
or black holes as remnants depending on the initial mass.

\begin{figure}
\resizebox{0.98\hsize}{!}{\includegraphics{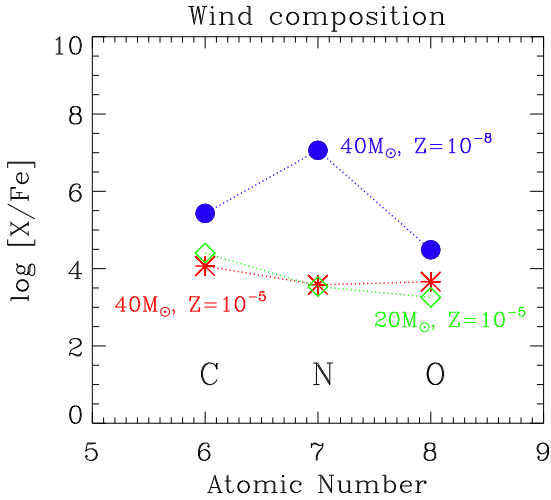}}
\caption{
CNO elements yields by stellar winds from 3 different evolutionary sequences
with $M_\mathrm{init} = 40$ and $Z=10^{-8}$ (filled circles), 
 $M_\mathrm{init} = 40$ and $Z=10^{-5}$ (asterisks), and  
 $M_\mathrm{init} = 20$ and $Z=10^{-5}$ (open squares). 
}
\label{fig:yields}
\end{figure}

\begin{figure}
\resizebox{0.98\hsize}{!}{\includegraphics{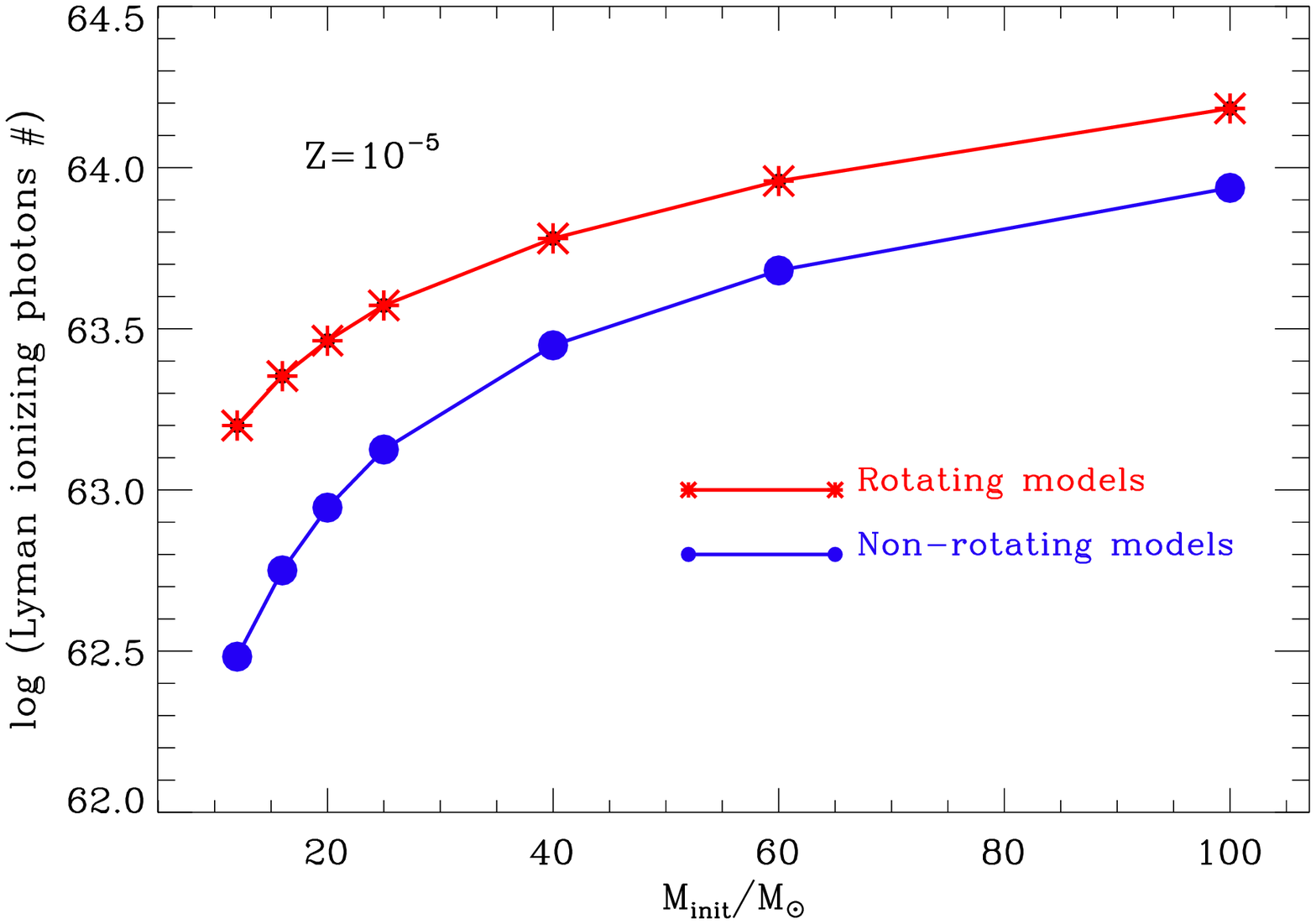}}
\resizebox{0.98\hsize}{!}{\includegraphics{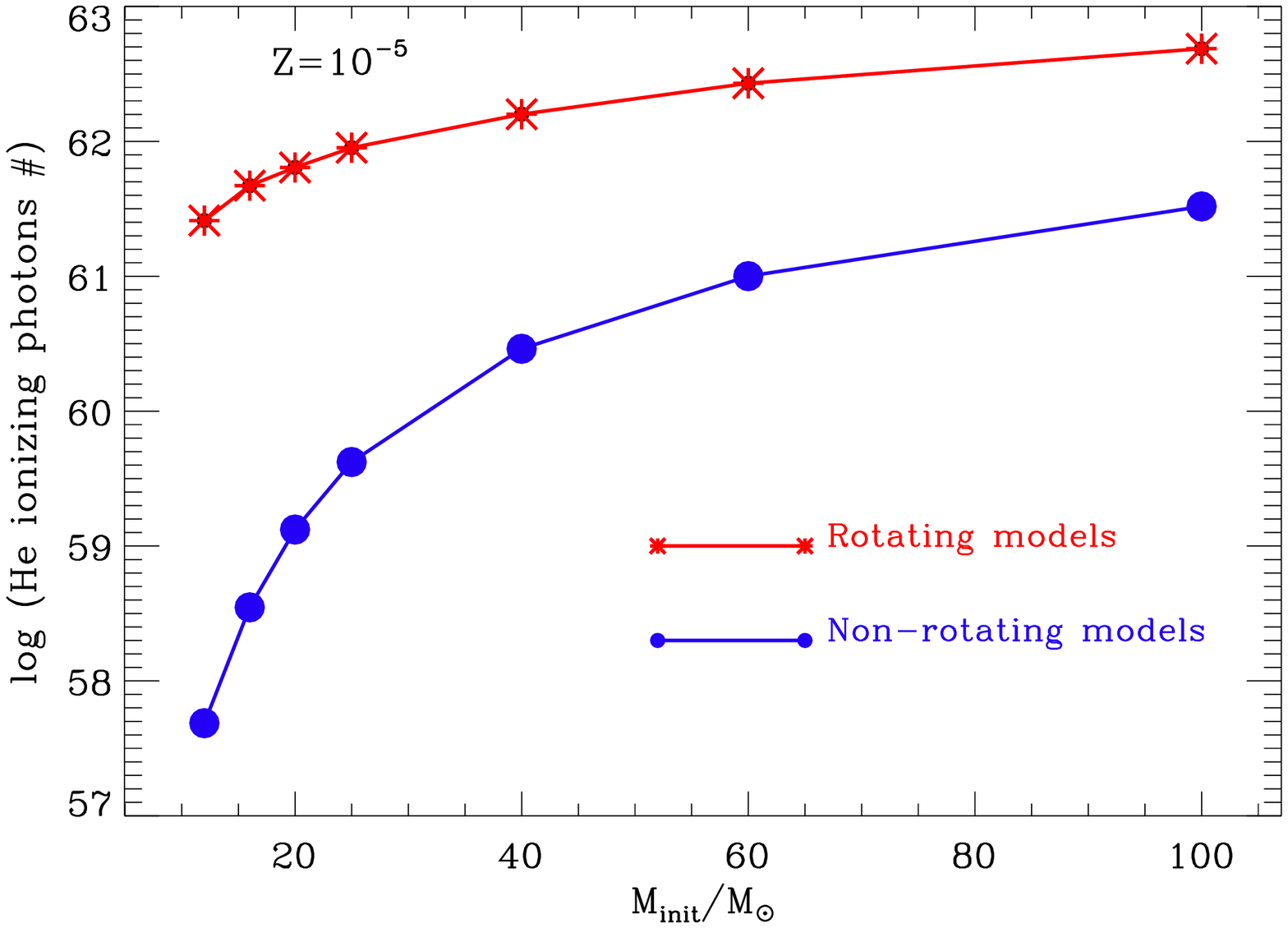}}
\caption{Total number of hydrogen (left panel) and helium (right panel) ionizing photons emitted throughout
the evolution of massive star models at $Z=10^{-5}$, 
as a function of the initial mass.
The lines connecting asterisks and filled triangles give the results from 
rapidly rotating stellar models that undergo the quasi-chemically
homogeneous evolution (see the text), and non rotating models, respectively.
}
\label{fig:photons}
\end{figure}

On the other hand, the quasi-chemically homogeneous evolution of rapidly rotating stars
have several interesting consequences. Avoiding the giant phase, some of them can retain 
a large amount of angular momentum in the core ($j_\mathrm{CO} \ge \sim 10^{16}~\mathrm{cm^2~s^{-1}}$) 
that may lead to production of a long gamma-ray burst (GRB) or a jet-induced energetic supernova/hypernova 
(Yoon \& Langer~2005; Woosley \& Heger~2006). 
A higher ratio of GRBs/HNe to SNe is predicted for lower metallicity (Yoon, Langer \& Norman~2006), 
and the role of the GRBs/HNe feedback may be significant in the early universe. 
E.g., such energetic supernovae might have left unique nucleosynthetic signatures in the early universe as discussed
by Nomoto (in this volume; Tominaga, Umeda \& Nomoto~2007), and could have affected the history of star formation at 
high redshift (Kobayashi, Spiringel \& White 2007). 
The very efficient rotationally induced chemical mixing also leads to abundant production of primary nitrogen 
as shown in Fig.~\ref{fig:yields}.  It is also remarkable that massive WR stars could be produced 
even at very low/zero metallicity due to the quasi-chemically homogeneous evolution, without the need of 
stellar winds. 
Such massive helum stars produce a few times more hydrogen ionizing photons and about 100 times 
more helium ionizing photons than the corresponding 
slowly rotating stars (Fig.~\ref{fig:photons}). 
This might have consequences in the history of reionization in the early universe, depending
on the initial rotational velocity function of the first and second generations of stars.  

\section{Evolution of binary stars at very low metallicity}

\begin{figure}
\resizebox{0.98\hsize}{!}{\includegraphics{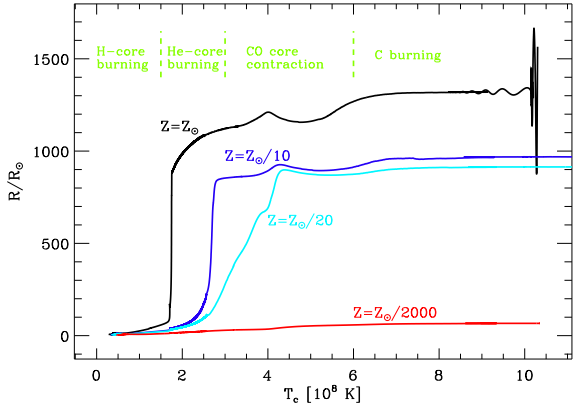}}
\caption{Change in the radii of $20~\mathrm{M_\odot}$ models with $V_\mathrm{rot, init} = 0.3 V_\mathrm{Kepler}$
at four different metallicities, as a function of central temperature. 
}
\label{fig:radius}
\end{figure}

Massive star evolution becomes much complicated by binary interactions
(e.g. de Greve \& de Loore~1992;  Podsiadlowski, Joss \& Hsu~1992; Pols~1995; 
Wellstein \& Langer~1999; Wellstein, Langer \& Braun~2001).
Binary star evolution is closely related to the change in the radii
of the stellar components throughout the evolutionary stages.
At solar metallicity, the envelope of a massive star drastically expands 
during the helium core contraction, 
and the case B mass transfer is most likely. 
As metallicity decreases, the rapid expansion of the envelope
of a massive star is delayed to a later evolutionary stage 
as revealed in Fig.~\ref{fig:radius}, and the Case C mass transfer
would become more common (see  de Mink et al. in this volume). 

In the stellar models at very low metallicity ($Z \le 10^{-5}$), 
massive stars remain fairly compact with the maximum radius less than a few hundreds solar radii throughout the evolution
if overshooting is ignored. 
This implies that a very close orbit may be required for a very metal poor star
to undergo Roche-lobe overflows in a binary system.
Our models also indicate that, at very low metallicity,  
mass transfer is usually dynamically stable regardless of the evolutionary stages of the primary, 
since the envelope remains radiative. 
However, rapid expansion of the envelope upto $\sim 10^3 \mathrm{R_\odot}$ 
during the CO core contraction
is observed in the models with $Z = 10^{-5}$ when overshooting is considered (de Mink et al. in this volume), 
and it remains uncertain whether very metal poor stars would 
become super-giants (see also Woosley in this volume).

A high mass transfer rate in a massive close binary usually leads
to thermal expansion of the envelope of the mass-accreting star. 
This effect becomes, however, less significant for lower metallicity 
and binary systems can avoid contact or common envelope phase more easily (de Mink et al., in this volume).
Therefore, it is expected that formation of compact binary systems such as black hole X-ray binaries via
common envelope phases should be less common at lower metallicity. 

\begin{figure}
\resizebox{0.99\hsize}{!}{\includegraphics{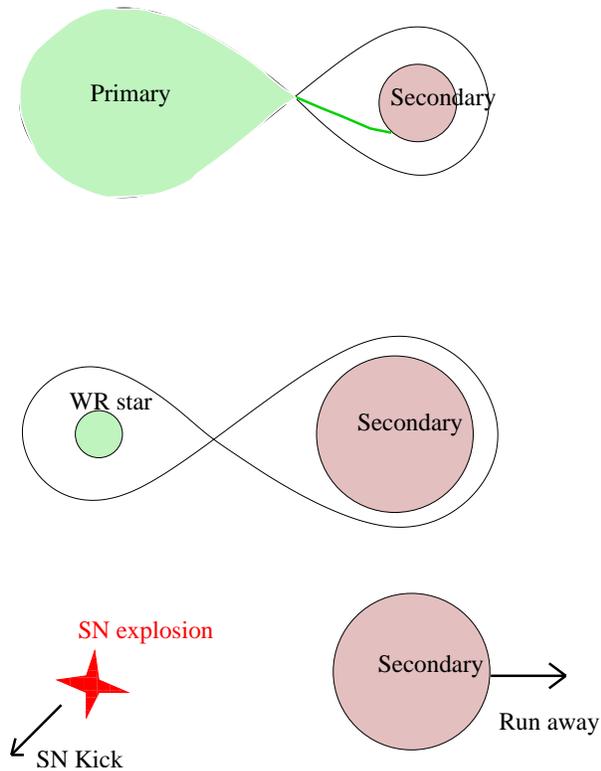}}
\caption{Runaway star scenario for long GRB progenitors. In a close metal poor binary system, 
the secondary gains mass and angular momentum from the primary. 
If the secondary is rejuvenated and spun up sufficiently as a result, 
it undergoes the quasi-chemically homogeneous evolution. 
The primary becomes a WR star and explode as a type Ib/c supernova. 
If the supernova kick unbounds the system, the secondary runs away at a velocity of $\sim 10...100~\mathrm{km/s}$, 
and explode as a long GRB/hypernova after traveling a few to several hundreds PCs away. 
See Cantiello et al.~(2007) for detailed discussions. 
}
\label{fig:binary}
\end{figure}

The role of rotation could be enhanced by binary interaction.
A slowly rotating star could be spun up to the break-up velocity by mass accretion 
(Langer et al.~2003; Petrovic, Langer \& van der Hucht~2005).
Models with fast semi-convection show that rapid mass accretion leads to rejuvenation
in mass-accreting stars, which weakens the chemical gradient between the core and the envelope
(Braun \& Langer~1995).
All these effects of mass accretion, i.e., 
increase in mass and angular momentum and rejuvenation, favor
chemical mixing by Eddington Sweet circulations in mass-accreting stars. 
Therefore, such a rapidly rotating star produced by mass accretion 
should experience strong chemical mixing during the rest of its life,
unless the orbit remains close enough for the star to be spun down 
by tidal interaction on a short time scale, after the mass transfer phase. 
If the secondary is rejuvenated sufficiently, even the quasi-chemically homogeneous evolution
can be induced by mass accretion, eventually leading to production of a long GRB
(Cantiello et al. 2007). 
This binary star scenario for long GRB progenitors 
suggests that a significant fraction of long GRBs should occur in runaway stars
(see Fig.~\ref{fig:binary} and Cantiello et al.~2007)
as implied by a recent observational study of Hammer et al (2006). 

In short, a large fraction of massive binaries at very low metallicity may undergo dynamically stable Case C mass transfer. 
Therefore, most primaries in close binaries would explode as type Ibc supernovae, 
while most secondaries as type II supernovae, but some of them as long GRBs/hypernovae.
The effects of chemical mixing such as production of primary nitrogen may
be significant in mass accreting secondary stars. 

\section{Concluding remarks}

We still do not fully understand the details
about the effects of rotation such as transport processes of 
angular momentum and chemical species
due to magnetic torques, Eddington Sweet circulations and other 
possible rotationally induced hydrodynamic instabilities in stars, 
which are important ingredients in recent stellar evolution models. 
Improved treatments of the rotation-related physics are thus needed
for future studies, in connection with observational tests and multi-dimensional
simulations
(e.g. Brott et al. in this volume; Talon~2007). 
However, above discussions clearly indicate that 
the evolution of metal poor massive stars 
much depends on rotation.
Addressing the role of the massive star feed back in the early universe 
including the effects of rotation and binarity will be
an exciting but challenging subject for the next decade.
In particular, future observational studies
on massive star populations (e.g. WR/O ratio) in metal poor galaxies
and supernovae/GRBs at high redshift
may provide strong constraints for theoretical models, 
as discussed in Yoon, Langer \& Norman~(2006) in greater detail. 




\begin{thebibliography}{}
\bibitem[\protect\citeauthoryear{Braithwaite}{2006}]{Braithwaite06}
Braithwaite, J., 2006, A\&A, 449, 451
\bibitem[\protect\citeauthoryear{Braun \& Langer}{1995}]{Braun95}
Braun, H., \& Langer, N., 1995, A\&A, 297, 483
\bibitem[\protect\citeauthoryear{Cantiello et al.}{2007}]{Cantiello07}
Cantiello, M.,  Yoon, S.-C., Langer, N. \& Livio, M., 2007, A\&A, 465, L29-L33
\bibitem[\protect\citeauthoryear{Castor, Abbott \& Klein}{1975}]{Castor75}
Castor, J.I., Abbott, D.C., \& Klein, R.I., 1975, ApJ, 195, 157
\bibitem[\protect\citeauthoryear{Chiappini et al.}{2006}]{Chiappini06}
Chiappini, C., Hirschi, R., Meynet, G., Ekstr\"om, S., Maeder, A., \& Matteucci, F., 2006, A\&A, 449, 27
\bibitem[\protect\citeauthoryear{Denissenkov \& Pinsonneault}{2007}]{Denissenkov06}
Denissenkov, P.A., \& Pinsonneault, M., 2007, ApJ, 655, 1157
\bibitem[\protect\citeauthoryear{de Greve \& de Loore}{1992}]{deGreve92}
de Greve, J.P., \& de Loore, C., 1992, A\&AS, 96, 653
\bibitem[\protect\citeauthoryear{Hammer et al.}{2006}]{Hammer06}
Hammer, F., Flores, H., Schaerer, D., Dessauges-Zavadsky, M., Le Floch, E., \& Puech, M., 2006, A\&A, 454, 103
\bibitem[\protect\citeauthoryear{Heger \& Langer}{2000}]{Heger00a}
Heger, A., \& Langer, N., 2000, ApJ, 544, 1016
\bibitem[\protect\citeauthoryear{Heger, Langer \& Woosley}{2000}]{Heger00b}
Heger, A., Langer, N., \& Woosley, S.E., 2000, ApJ, 528, 368
\bibitem[\protect\citeauthoryear{Heger, Woosley \& Spruit}{2005}]{Heger05}
Heger, A., Woosley, S.E., \& Spruit, H.C., 2005, ApJ, 626, 350
\bibitem[\protect\citeauthoryear{Hirschi, Meynet \& Maeder}{2004}]{Hirschi04}
Hirschi, R., Meynet, G., \& Maeder, A., 2004, A\&A, 425, 649
\bibitem[\protect\citeauthoryear{Langer}{2007}]{Kobayashi07}
Kobayashi, C., Springel, V., \& White, S.D.M., 2007, MNRAS, 376, 1465
\bibitem[\protect\citeauthoryear{Langer}{1992}]{Langer92}
Langer, N., 1992, A\&A, 265, 17
\bibitem[\protect\citeauthoryear{Langer et al.}{1999}]{Langer99}
Langer, N., Heger, A., Wellstein, S., \& Herwig, F., 1999, A\&A, 346, 37
\bibitem[\protect\citeauthoryear{Langer et al.}{1999}]{Langer99}
Langer, N., Yoon, S.-C., Petrovic, J., \& Heger, A., 2003, in Stellar Rotation, Proc. IAU-Symp. 215, ASP, San Francisco, A. Maeder, P. Eenens, eds. 
\bibitem[\protect\citeauthoryear{Maeder}{1987}]{Maeder87}
Maeder, A., 1987, A\&A, 178, 159
\bibitem[\protect\citeauthoryear{Maeder \& Meynet}{2000}]{Maeder00}
Maeder, A., \& Meynet, G., 2000, ARA\&A, 38, 143
\bibitem[\protect\citeauthoryear{Maeder \& Meynet}{2005}]{Maeder05}
Maeder, A., \& Meynet, G., 2005, A\&A, 440, 1041
\bibitem[\protect\citeauthoryear{Meynet \& Maeder}{2003}]{Meynet03}
Meynet, G., \& Maeder, A., 2003, A\&A, 390, 561
\bibitem[\protect\citeauthoryear{Mokiem et al.}{2006}]{Mokiem06}
Mokiem, M.R., de Koter, A., \& Evans, C.J. et al., 2006, A\&A, 456, 1131
\bibitem[\protect\citeauthoryear{Mokiem et al.}{2007}]{Mokiem07}
Mokiem, M.R., de Koter, A., \& Evans, C.J. et al., 2007, A\&A, 465, 1003
\bibitem[\protect\citeauthoryear{Spruit}{2002}]{Spruit02}
Spruit, H.C., 2002, A\&A, 381, 923
\bibitem[\protect\citeauthoryear{Pauldrach, Puls \& Kudritzki}{1986}]{Pauldrach86}
Pauldrach, A.W.A., Puls, J., \& Kudritzki, R.P., 1986, A\&A, 164, 86
\bibitem[\protect\citeauthoryear{Petrovic, Langer \& van der Hucht}{2005}]{Petrovic05a}
Petrovic, J., Langer, N., \& van der Hucht, K.A., 2005, A\&A, 435, 1013
\bibitem[\protect\citeauthoryear{Petrovic et al.}{2005}]{Petrovic05b}
Petrivic, J., Langer, N., Yoon, S.-C., \& Heger, A., 2005, A\&A, 435, 247
\bibitem[\protect\citeauthoryear{Podsiadlowski}{1992}]{Podsiadlowski92}
Podsiadlowski, Ph., Joss, P.C., \& Hsu, J.J.L., 1992, ApJ, 391, 246
\bibitem[\protect\citeauthoryear{Pols}{1994}]{Pols94}
Pols, O.R., 1994, A\&A, 290, 119
\bibitem[\protect\citeauthoryear{Talon}{2007}]{Talon07}
Talon, S, 2007, [astro-ph/0708.1499]
\bibitem[\protect\citeauthoryear{Tominaga}{2007}]{Tominaga07}
Tominaga, N., Umeda, H., \& Nomoto, K., 2007, ApJ, 660, 516
\bibitem[\protect\citeauthoryear{Vink, de Koter, \& Lamers}{2001}]{Vink01}
Vink, J.S., de Koter, A., \& Lamers, H.J.G.L.M., 2001, A\&A, 369, 574
\bibitem[\protect\citeauthoryear{Wellstein \& Langer}{1999}]{Wellstein99}
Wellstein, S.,  \& Langer, N., 1999, A\&A, 350, 148
\bibitem[\protect\citeauthoryear{Wellstein, Langer \& Braun}{2001}]{Wellstein01}
Wellstein, S.,  Langer, N., \& Braun, H.,  2001, A\&A, 369, 939
\bibitem[\protect\citeauthoryear{Woosley \& Heger}{2006}]{Woosley06}
Woosley, S.E., \& Heger, A., 2006, ApJ, 637, 914
\bibitem[\protect\citeauthoryear{Yoon \& Langer}{2005}]{Yoon05}
Yoon, S.-C., \& Langer, N., 2005, A\&A, 443, 643
\bibitem[\protect\citeauthoryear{Yoon, Langer \& Norman}{2006}]{Yoon06}
Yoon, S.-C., Langer, N., \& Norman, C.,  2006, A\&A, 460, 199
\bibitem[\protect\citeauthoryear{Zahn}{1992}]{Zahn92}
Zahn, J.-P., 1992, A\&A, 265, 115
\bibitem[\protect\citeauthoryear{Zahn, Brun \& Mathis}{2007}]{Zahn07}
Zahn, J.-P., Brun, A.S., \& Mathis, S., 2007, A\&A, in press, [astro-ph/0707.3287]

\end{thebibliography}

\end{document}